\documentclass[
aip,
amsmath,amssymb, 
reprint
]{revtex4-1}

\usepackage{graphicx}
\usepackage{mathptmx}
\usepackage{hyperref}
\usepackage[utf8]{inputenc}
\usepackage[T1]{fontenc}
\usepackage{etoolbox}
\DeclareUnicodeCharacter{0308}{\"u}

\begin{document}

\title{Observation of sequential three-body dissociation of camphor molecule - a native frame approach}
\vspace{10 mm}
\author{S. De}
\affiliation{Quantum Center of Excellence for Diamond and Emergent Materials and Department of Physics, Indian Institute of Technology Madras, Chennai 600036, India.}

\author{S. Mandal}
\affiliation{Quantum Center of Excellence for Diamond and Emergent Materials and Department of Physics, Indian Institute of Technology Madras, Chennai 600036, India.}
\affiliation{Indian Institute of Science Education and Research, Pune~411008, India.}

\author{Sanket Sen}
\affiliation{Indian Institute of Technology Hyderabad, Kandi~502285, India.}

\author{Arnab Sen}
\affiliation{Indian Institute of Science Education and Research, Pune~411008, India.}

\author{R. Gopal}
\affiliation{Tata Institute of Fundamental Research, Hyderabad~500046, India.}

\author{L. Ben Ltaief}
\affiliation{Department of Physics and Astronomy, Aarhus University, 8000 Aarhus C, Denmark}

\author{S. Turchini}
\affiliation{Istituto di Struttura della Materia - CNR (ISM-CNR), Area di Ricerca di Tor Vergata via del Fosso del Cavaliere, 100, 00133 Rome, Italy.}

\author{D. Catone}
\affiliation{Istituto di Struttura della Materia - CNR (ISM-CNR), Area di Ricerca di Tor Vergata via del Fosso del Cavaliere, 100, 00133 Rome, Italy.}

\author{N. Zema}
\affiliation{Istituto di Struttura della Materia - CNR (ISM-CNR), Area di Ricerca di Tor Vergata via del Fosso del Cavaliere, 100, 00133 Rome, Italy.}

\author{M. Coreno}
\affiliation{Elettra-Sincrotrone Trieste, 34149 Basovizza, Italy.}
\affiliation{Istituto di Struttura della Materia - CNR (ISM-CNR), Trieste branch,in Basovizza Area Science Park, 34149 Trieste,  Italy.}

\author{R. Richter}
\affiliation{Elettra-Sincrotrone Trieste, 34149 Basovizza, Italy.}

\author{M. Mudrich}
\affiliation{Quantum Center of Excellence for Diamond and Emergent Materials and Department of Physics, Indian Institute of Technology Madras, Chennai 600036, India.}
\affiliation{Department of Physics and Astronomy, Aarhus University, 8000 Aarhus C, Denmark}

\author{V. Sharma}
\email{vsharma@phy.iith.ac.in}
\affiliation{Indian Institute of Technology Hyderabad, Kandi~502285, India.}

\author{S. R. Krishnan}
\email{srkrishnan@iitm.ac.in}
\affiliation{Quantum Center of Excellence for Diamond and Emergent Materials and Department of Physics, Indian Institute of Technology Madras, Chennai 600036, India.}

\begin{abstract} 
The three-body dissociation dynamics of the dicationic camphor molecule (C$_{10}$H$_{16}$O$^{2+}$) resulting from Auger decay are investigated using soft X-ray synchrotron radiation. 
A photoelectron-photoion-photoion coincidence (PEPIPICO) method, a combination of a velocity map imaging (VMI) spectrometer and a time-of-flight (ToF) spectrometer is employed to measure the $3$D momenta of ions detected in coincidence. 
The ion mass spectra and the ion-ion coincidence map at photon energies of $287.9$~eV (below the C~$1$\textit{s} ionization potential) and $292.4$~eV (above the C~$1$\textit{s} ionization potential for skeletal carbon) reveal that fragmentation depends on the final dicationic state rather than the initial excitation. 
Using the native frame method, three new fragmentation channels are discussed; ($1$) CH$_2$CO$^+$ + C$_7$H$_{11}^+$ + CH$_3$, ($2$) CH$_3^+$ + C$_7$H$_{11}^+$ + CH$_2$CO, and ($3$) C$_2$H$_5^+$ + C$_6$H$_9^+$ + CH$_2$CO. 
The dominating nature of sequential decay with deferred charge separation is clearly evidenced in all three channels. 
The results are discussed based on the experimental angular distributions and momenta distributions, corroborated by geometry optimization of the ground, monocationic, and dicationic camphor molecule.
\end{abstract}

%

\vspace{2 mm}

%
\maketitle
%
%


\section{Introduction}
A key component of chemical reactions and spectroscopic investigations is the fragmentation of molecules, which provides essential insights into the intricate details of molecular structures and their dynamic behaviour.
A study of molecular fragmentation dynamics is necessary to comprehend fundamental chemical processes, including radiation damage to biological systems~\cite{de_vries_charge_2003, markush_role_2016, kc_towards_2022, schlatholter_ion-induced_2006} and the stability of systems that are crucial for astrophysics~\cite{wei_isomerization_2021, zhen_quadrupole_2014}. 
Among the different ways of investigating these dynamics, the detection of electrons produced by Auger decay~\cite{hikosaka_auger_2006, carniato_single_2020} is instrumental in providing a wealth of information about the electronic rearrangements and energy redistributions accompanying molecule fragmentation. 
When an electron from the core-shell of an atomic or molecular system is excited or ionized, a subsequent electron from the outer shell fills the resultant core hole.
The excess energy resulting from this electronic relaxation is then used to either emit a photon or remove single or multiple electrons from the system, which may lead to multiple ionizations. The formation of a highly charged state via Auger decay is often followed by fragmentation of molecular systems.
Auger decay deposits a large amount of energy in the system, which is redistributed into different degrees of freedom and nuclear motion of the system~\cite{travnikova_ultrafast_2022, Li_ultrafast_2015}. 
The deposited internal energy~\cite{kukk_internal_2015, saha_state_2014, levola_ionization-site_2015, itala_photofragmentation_2016, salen_resonant_2020, pihlava_photodissociation_2021, ganguly_coincidence_2022} and its redistribution~\cite{erdmann_general_2021, xie_insights_2020} has a significant impact on the dissociation dynamics of the system.
Additionally, ultrafast charge migration is also reported by Erk \emph{et al.}~\cite{erk_ultrafast_2013} in inner shell ionization of CH$_3$SeH, where the charge state of Se is smaller in the molecular environment as compared to the isolated atom. 
The dissociation path also depends on the existence of ultrafast restructuring of atomic components, such as hydrogen/proton migration~\cite{hishikawa_hydrogen_2004, xu_tracing_2009, Zhang_2012}. 
Furthermore, the fragmentation pattern differs based on whether the C 1\textit{s} electron promoted into a Rydberg orbital or an antibonding $\pi$* molecular orbital~\cite{eberhardt_site_specific_1983}.
Auger electron-photoion-photoion coincidence investigations demonstrated the crucial significance of the internal energy on the fragmentation dynamics in the case of CH$_2$ClBr, where strong site-specific fragmentation was found at low internal energies~\cite{miron_site_1998}.
However, Inhester \emph{ et al.}~\cite{Inhester_site_2021} demonstrated that the fragmentation pattern depends on the final dicationic two-hole state formed by Auger decay rather than the internal energy distribution. They also observed low site specificity on the molecular dissociation of ethyl trifluoroacetate.
A prior investigation of OCS revealed that the final state achieved, following Auger decay, significantly affects the fragmentation mechanisms~\cite{saha_state_2014}.\par
In large molecules~\cite{abid_electronion_2020}, many-body fragmentation channels are known to happen. In the context of a large molecule, camphor is a naturally occurring volatile, monoterpene ketone known for its medicinal and other uses\cite{gaddam_recent_2015, zielinska-blajet_monoterpenes_2020}.
This system has long served as a benchmark chiral molecule for asymmetries in angle-resolved photoelectron spectra~\cite{nahon_determination_2006,nahon_effects_2010,lux_circular_2012, Lux_circular_2015, garcia_circular_2003,devlin_ab_1997,blanchet_ultrafast_2021,hergenhahn_photoelectron_2004}. Although the study does not delve into the chirality of the molecular system, it is important to note that camphor molecules undergo intricate bond rearrangements during fragmentation due to their complexity and numerous functional groups~\cite{weinberg_mass_1966, dimmel_1967_preferential}. Understanding how various bonds break and reorganize during the fragmentation process can enhance our comprehension of the complex mechanics of such photochemical reactions.

The development of simultaneous detection of photoelectrons and photoions in coincidence (PEPICO, PEPIPICO) allows a deeper understanding of molecular fragmentation dynamics. 
PEPIPICO has been used in a number of recent studies to uncover the kinetics of molecular fragmentation of both organic~\cite{Ha_photofragmentation_2011} and inorganic~\cite{lindgren_molecular_2005} molecular dicationic states. 
In our previous study~\cite{sen_fragmentation_2022}, we demonstrated experimental evidence of four fragmentation channels. 
Two experimentally observed fragmentation channels of camphor, (\textbf{i}) CH$_3^+$ + C$_8$H$_{13}^+$ + CO, and (\textbf{ii}) C$_4$H$_7^+$ + C$_5$H$_9^+$ + CO, were corroborated using molecular dynamics (MD) simulations at two different bath temperatures $2500$~ K and $4000$~ K, (which correlates with the internal energy deposited in the molecule following Auger decay).
At bath temperature of $2500$~K, the dicationic molecule follows a deferred charge separation decay with the emission of a neutral CO (m/q $28$) in the $1$st step, followed by fragmenting to CH$_3^+$ (m/q $15$ ) \& C$_8$H$_{13}^+$ (m/q $109$) in the $2$nd step.
Up to $4000$~K, this fragmentation channel remains active. Between $4000$~K and $5000$~K, the molecule dissociates and emits neutral CO (m/q $28$) in the $1$st step of the deferred charge separation process, followed by the emission of C$_4$H$_7^+$ (m/q $55$) \& C$_5$H$_9^+$ (m/q $69$) fragment ions in the $2$nd step.
Notably, the experimentally observed energy distribution of neutral CO, which was released in deferred charge separation dissociation, showed the signature of bath temperature in MD simulations. 

In this paper, we used an oppositely positioned Time-of-Flight (ToF) spectrometer and velocity map imaging (VMI) spectrometer for photoelectron-photoion-photoion coincidence (PEPIPICO) detection, a VMI-PEPIPICO system. 
We extend our discussion on the fragmentation of camphor molecules following excitation near C~$1$\textit{s} potential and report three new fragmentation channels for the first time. 
The photon energies of $287.9$~eV and $292.4$~eV were used to excite and ionize the C~$1$\textit{s} core electron of the camphor molecule, respectively. The recorded PEPIPICO maps clearly show similar fragmentation for both photon energies. 
This indicates that the fragmentation depends on the final dicationic state rather than the initial excitation mechanism. 
From the experimentally observed angular and momenta distributions, three deferred charge separation pathways, ($1$) CH$_2$CO$^+$ + C$_7$H$_{11}^+$ + CH$_3$, ($2$) CH$_3^+$ + C$_7$H$_{11}^+$ + CH$_2$CO, and ($3$) C$_2$H$_5^+$ + C$_6$H$_9^+$ + CH$_2$CO along with one concerted dissociation channel C$_2$H$_5^+$ + C$_6$H$_9^+$ + CH$_2$CO are identified. 
We have observed that in the second step of sequential dissociation, the intermediate metastable ion dissociates into two charged fragments via Coulomb repulsion, similar to our previous study~\cite{sen_fragmentation_2022}. In contrast, when heavy (charged) particles collide, they are highly effective at exciting molecules to reach highly charged states.
This can cause the molecules to "explode" due to the intense internal Coulomb forces, known as ``Coulomb explosion''. The Coulomb explosion channel generally follows a repulsive curve. Had that dissociation been through the Coulomb explosion, the fragment ions would have carried very high energy. However, when molecules interact with synchrotron radiation or electron impact, they become ionized to lower charged states and dissociate via Coulomb repulsion. So, the fragmentation channels considered for the study do not arise from the Coulomb explosion. These observations underscore the need for comprehensive theoretical investigations to gain a profound understanding of the intricate processes involving relaxation and dissociation of the dicationic camphor molecule.

\section{Materials and Methods}
\subsection{Experimental Methods}
\begin{figure}
\includegraphics[width= \linewidth]{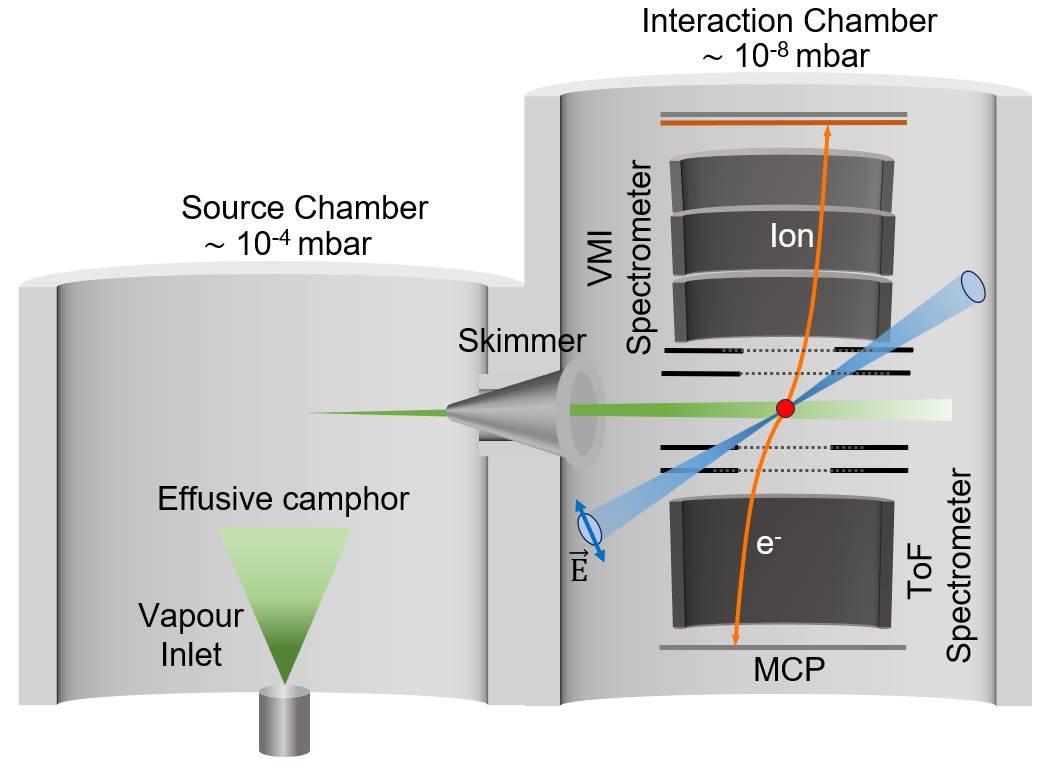}
\caption{
Schematic diagram of the experimental setup: Vaporized camphor molecules effuse into the source chamber through a dosing valve. camphor molecules enter the interaction chamber, where a VMI spectrometer is housed in the top half, and a ToF spectrometer is housed in the bottom. Ions are detected by the VMI spectrometer with the coincidence detection of electrons by the ToF spectrometer. Electron-ion (single) and electron-ion-ion (double) coincidences are detected by using multi-coincidence electronics.
}
\label{figure: Experimental_setup}
\end{figure}

The experiments were carried out on the Circular Polarization (CiPo) beamline~\cite{derossi_high_1995} of Elettra synchrotron, Trieste, Italy. 
An extended description of a similar experimental setup can be found in \cite{mandal_penning_2020, buchta_charge_2013}. 
A schematic of the experimental setup is presented in Figure~\ref{figure: Experimental_setup}.
Briefly, the setup can be divided into two parts: the source chamber, where camphor vapour was effused by a controlled leak through a dosing valve, maintained at a chamber pressure of $10^{-4}$~mbar. 
High purity ($> 95 \% $) ($1$\textit{S})-($-$)-Camphor from Sigma-Aldrich was used without further purification. 
Then, camphor molecules enter the second part of the experimental setup, the interaction chamber, which is maintained at a lower pressure of $10^{-8}$~ mbar, through a skimmer. 
Here, camphor molecules interact with the soft X-ray synchrotron photons. A VMI spectrometer was used to record the ions coincident with the electrons detected by a ToF spectrometer mounted oppositely, working in tandem. 
Thus, electrons are detected without energy or angle resolution, while the VMI is used to detect ions in all the data presented in this work. 
This allowed us to develop the complete momenta vector of the photoions. Synchronous use of VMI and ToF made it possible to record electron - multi-ion coincidence measurements, which is used to distinguish different fragmentation channels further. 
The ion imaging energy resolution is $\Delta{E}/E \sim 20 - 9\%$ for ion kinetic energies between $0.2$~eV and $5.0$~eV. The energy of photons provided by this beamline is in the range of $5 - 1200$~eV, which well includes our desired photon energy range, $280 - 305$~eV. 
The Spherical Grating Monochromator, G$3$, which is intended to operate in the $120 - 400$~eV range, showed an effective resolving power of $\sim 4 \times 10^3$ for the photon energies used in our experiments. 
The photon energy was calibrated using C~$1$s $\rightarrow \pi$* excitation of CO$_2$~\cite{sen_fragmentation_2022}

From the recorded total ion yield spectra~\cite{sen_fragmentation_2022}, we can observe that the primary ionization cross-section becomes a maximum at $292.4$~eV with a secondary peak around $287.9$~eV. Therefore, to get a better signal-to-noise ratio, we decided to investigate the fragmentation of camphor molecules at these photon energies.
Further, as the photon energy $292.4$~eV, lies in between the C~$1$\textit{s} ionization potential of the carbonyl C  ($293$~eV) and other skeleton C ($290.5$~eV), normal Auger decay will be the dominant process for the skeleton C, whereas for the carbonyl C, only resonant Auger decay will occur. 
In contrast, at $287.9$~eV, due to the excitation of the core electrons, only resonant Auger decay will occur.
The collection efficiency of our system is higher for low kinetic energy electrons. Therefore, the fast resonant Auger electrons are mostly missed.
Thus, the overall collection efficiency of the electrons produced at $287.9$~eV will be less than those produced at a photon energy of $292.4$~eV.
 
\subsection{Analysis techniques}
The complete $3$D momenta ($p_x$, $p_y$, $p_z$) information of the fragment ions can be obtained from the position information ($X$ and $Y$ coordinate on the delay line anode of VMI) and the time of flight ($T$) information from each ionization event. Post-ionization detection of an electron by ToF spectrometer, a "start" signal is generated, which is considered a ``time zero'' reference. The flight time ($T$) of the ionic fragments is measured with respect to that ``time zero'' reference. Then, the initial momenta components of the ionic fragment can be calculated as follows~\cite{gopal_note_2018, khan_recoil_2015, sen_fragmentation_2022}:
\begin{equation}
   p_x = \frac{mX}{fT}
\end{equation} \par
\begin{equation}
   p_y = \frac{mY}{fT}
\end{equation} \par
\begin{equation}
    p_z = -CqE_s\Delta{T}
\end{equation}
\par
Here $m$ and $q$ are the mass and the charge of the ionic fragment, respectively, and $E_s$ stands for the electric field in the extraction region.
$\Delta{T}$ denotes the ToF difference between the fragment ion with an initial momentum along the spectrometer axis ($p_z \not= 0$) and the fragment ion of the same mass and charge but with a zero initial momentum component along the spectrometer axis ($p_z = 0$).
The parameters $f$ and $C$ correspond to the correction factors of the lensing action due to the non-uniform field, which affect the path of the ions and, thus, the flight time. 
These two factors have been calculated using SIMION 8.0 (Scientific Instrument Services, Inc., USA); a detailed discussion can be found in \cite{gopal_note_2018}.
Furthermore, for a three-body fragmentation: ABC$^{2+} \rightarrow$ A$^++~$B$^+ +$ C, the neutral fragment's momentum
($\Vec{P}_{C}$)
can be calculated from the conservation of momentum, where 
\begin{equation}
{\Vec{P}_{C}} = -{\left(\Vec{P}_A + \Vec{P}_B\right)}
\end{equation} \par
where $\Vec{P}_A$ and $\Vec{P}_B$ are the initial momenta of the first and second fragment ion, respectively. 

Representation of different dissociation channels in the form of Dalitz plot~\cite{bittner_sequential_2022, wales2014coulomb, bhattacharyya_two-_2022, yan_observation_2016, khan_observation_2015, Shen_fragmentation_2016} and the Newton diagram~\cite{hsieh_reaction_1997, Neumann_fragmentation_2010, bapat_bent_2007, abid2023hydrogen, kling2019time, yuan_three-body_2022, wu_nonsequential_2013, bhattacharyya_two-_2022, yan_observation_2016} are particularly helpful in determining the dissociation process. 
However, a complete separation of sequential and concerted decay mechanisms is not possible by either of these data visualization techniques.
Rajput \emph{et al.}~\cite{Rajput_native_2018} developed a novel method called "native frame"~\cite{bhattacharyya_two-_2022, yadav_hydrogen_2022, McManus_disentangling_2022, Rajput_addressing_2023} analysis to distinguish sequential and concerted fragmentation channels in a more complete way. 
The idea of this new technique is to assume initially that the dissociation is occurring sequentially in a two-step process and then determine the conjugate momenta at each step of dissociation. 
If, for the intermediate molecular ion, the lifetime of the electronic state exceeds its rotational time period, then the angular correlation between the $1$st step and the $2$nd step of dissociation is lost. Therefore, the angular distribution of the $2$nd step with respect to the $1$st step of dissociation is intuitively expected to be uniform for these electronic states.
To visualize this, one can do a $3$D plot between the KER release in the $2$nd step of dissociation and the angular distribution of the $2$nd step of dissociation with respect to the $1$st step of dissociation, where the third dimension shows the counts.
Therefore, sequential dissociation is identified as the uniform distribution in the $3$D density plot. In contrast, the concerted dissociation forms a localized concentrated region in the density plot.
This characteristic signature of the sequential and concerted breakups helps to identify the dissociation dynamics, and therefore, one can selectively subtract sequential breakup events to get the branching ratio of concerted and sequential breakups. 
In native frames, initially, it is assumed that the dissociation occurs in a two-step process via the formation of a metastable intermediate ion: ABC$^{2+} \rightarrow $ AB$^{2+} + $ C $\rightarrow$ A$^+ + $B$^+ +$ C. The relative angle between the $2$nd step of dissociation in center-of-mass reference frame of AB$^{2+}$ and the $1$st step of dissociation in center-of-mass reference frame of ABC$^{2+}$ is given by 
\begin{equation}
    \theta_{AB^{2+}-C} = {\cos^{-1} \left[\frac{\Vec{P}_{1st}.\Vec{P}_{2nd}}{\mid \Vec{P}_{1st} \mid \mid \Vec{P}_{2nd}\mid }\right]}
\end{equation}
Here, $\Vec{P}_{1st}$  and $\Vec{P}_{2nd}$ are the conjugate momenta associated with the $1$st step and $2$nd step of the dissociation process, respectively.
\begin{equation}
    \Vec{P}_{1st} = \frac{m_{AB}}{M}\Vec{P_C} - \frac{m_{c}}{M}(\Vec{P_A} + \Vec{P_B})
\end{equation}
\begin{equation}
    \Vec{P}_{2nd} = m_{AB}\left(\frac{\Vec{P_B}}{m_B} - \frac{\Vec{P_A}}{m_A}\right)
\end{equation}

Hence, the kinetic energy release (KER$_{AB^{2+}}$) in the $2$nd step of the dissociation process is given by
\begin{equation}
    \mathrm{KER}_{AB^{2+}} = \frac{\Vec{P}_{2nd}{^2}}{2m_{AB}}
\end{equation}
Here, m$_{AB}$ is the reduced mass of the intermediate AB$^{2+}$ ion.

\begin{figure*}[ht]
\centering
\includegraphics[width= 15 cm]{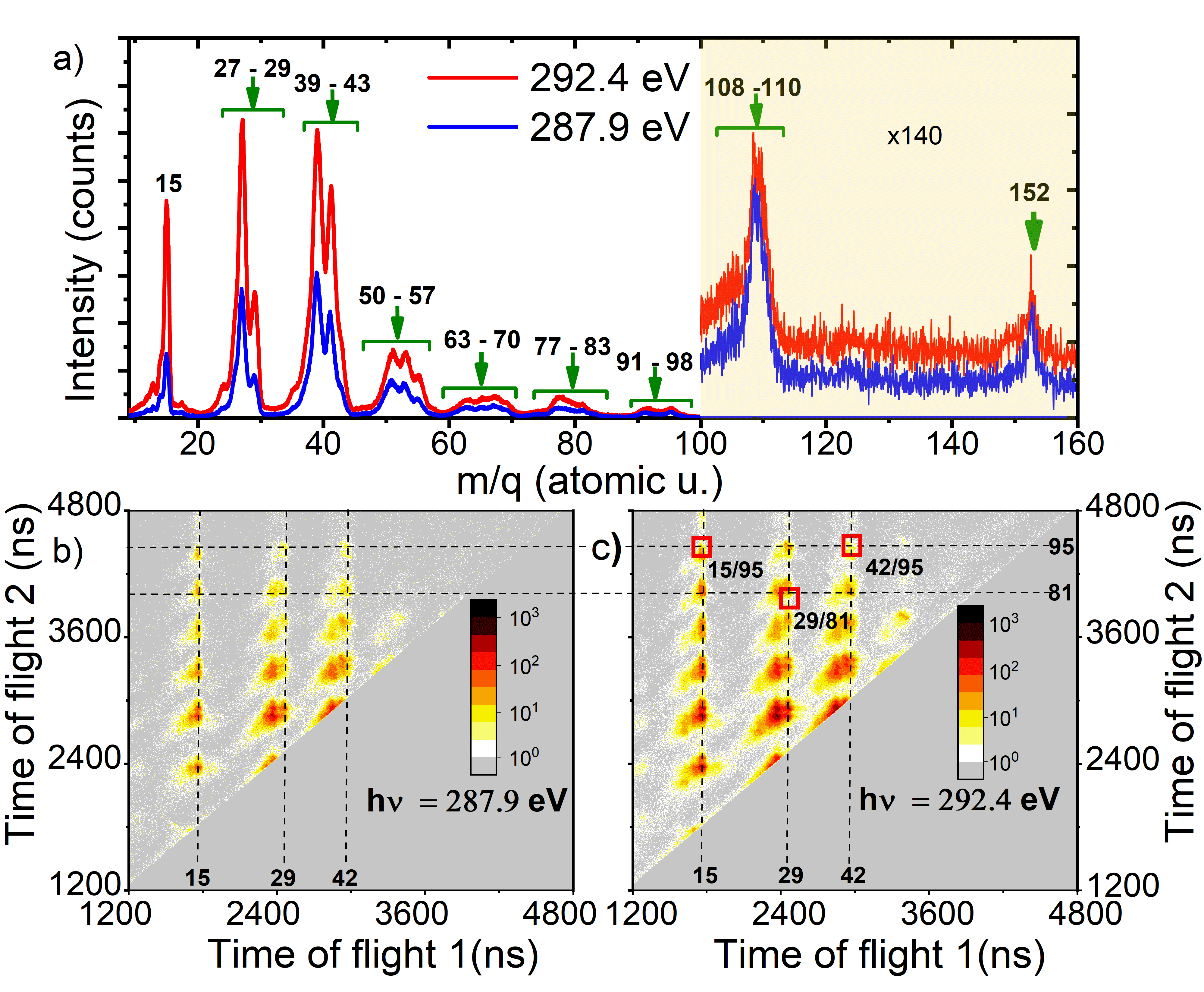}
\caption{\text{a}) Comparison of the ion mass spectra for photoionization at h$\nu= 287.9$~eV (blue spectrum) with that at $292.4$~eV of Sen \textit{et al.} (red spectrum), cf. Figure 5 a) of ref. \cite{sen_fragmentation_2022}, up to mass-to-charge of 160 atomic unit. The two mass spectra are plotted on the same scales for suitable comparison. The abscissa represents the mass-to-charge ratio of fragment ions, and the ordinate represents the intensity on a linear scale. The shaded region (light yellow colour) represents $140$ times magnification of the intensity. Further, \text{b}) and \text{c}) represent the comparison of coincidence maps at photon energies of $287.9$~eV with that at $292.4$~eV of Sen \textit{et al.} (red spectrum), cf. Figure 5 b) of ref. \cite{sen_fragmentation_2022}, respectively, where time of flight $1$ is the arrival time of the faster ion and time of flight $2$ is the arrival time of the slower ion. The horizontal and vertical dashed lines show the correlated ion mass for some selected islands marked by red rectangular boxes.
}
\label{figure: Mass_spectra_coincidence_map}
\end{figure*} 
\section{Results and discussions}
\subsection{Mass spectra and ion-ion coincidence map}
The typical ion mass spectra recorded in the PEPIPICO mode at two photon energies (\textbf{i}) $287.9$~eV (below the C~$1$\textit{s} ionization potential of the camphor molecule), and (\textbf{ii}) $292.4$~eV\cite{sen_fragmentation_2022} (above the C~$1$\textit{s} ionization potential of the skeletal carbon atoms of the camphor molecule) are shown in Figure~\ref{figure: Mass_spectra_coincidence_map}~\text{a}). The ion mass spectrum at $292.4$~eV was also presented in our recent work\cite{Sen_2024} up to mass-to-charge ratio $350$ atomic unit. Both the mass spectra at photon energies of $287.9$~eV and $292.4$~eV, respectively, contain mass peaks at $m/q = 15,~27-29,~39-43,~50-57,~63-70,~77-83,~91-98,~108-110,~$and$~152$. Compared to electron impact ionization studies~\cite{weinberg_mass_1966, nist_camphor}, the relative ratio of smaller to larger fragments formed near C~$1$\textit{s} ionization is different. This is due to the presence of highly efficient secondary ionization mechanisms, such as Auger decay~\cite{kivimaki_metastable_2014}, which are accessible in our case, cause the molecules to be doubly or triply ionized, leading to extensive fragmentation.
Identical fragment ions are observed with different intensities in the mass spectra of the camphor molecule at both the photon energies of our study. 
Castilho \textit{et al.}~\cite{de_castilho_single_2014} have also reported a similar observation of the varying intensity of different ionic fragments with the change in photon energies. Although their reported photoionization at $270$~eV differs from the decay at the photon energies we are studying here. Various possible mechanisms are possible in this energy range. All outer, inner valence, and double (and higher) continua are essentially open at $270$~eV, although C~$1$\textit{s} Auger decay would not happen. The continua are still open at $287.9$~eV, but a resonant Auger decay is feasible, resulting in a singly ionized one-hole or two-hole one-particle configuration, indicated in the shown mass spectra by more parent ions in the blue spectrum.
The states above the double ionization can decay further by cascades.
At the highest energy $292.4$~eV (above the first C~$1$\textit{s} ionization potential of all the skeleton C atoms), the normal Auger process becomes available (in addition to possible resonant Auger from states converging to the higher threshold).
The normal Auger process forms doubly ionized states, where the involved two electrons are the C~$1$\textit{s} photoelectron and the Auger electron. 
Additionally, fluorescence at lower energies subsequent to Auger decay can still occur.
Direct x-ray fluorescence decay, particularly from the C~$1$\textit{s} level, is expected to be a relatively minor pathway compared to ionization.
Thus, more double ionization would be expected above the threshold (this process is certainly preferred), reflected in the mass spectra shown by more pronounced fragment ion mass peaks (red spectrum) and also by the intense islands in the ion-ion coincidence map

The observed similarities in fragmentation at photon energies of $287.9$~eV and $292.4$~eV could possibly be attributed to the same final dicationic states achieved by the molecule before fragmentation, independent of initial excitation as suggested by Castilho \textit{et al}.~\cite{de_castilho_single_2014} Generally, the doubly ionized final states would have a very short lifetime and be strongly dissociative. Hazra \textit{et al.}\cite{hazra_vibronic_2005} has shown that for energetically closely spaced excited states, the Born-Oppenheimer approximation (B–O) is not valid. One consequence of the breakdown of the validity of the B-O approximation is surface crossings, or conical intersections, that might lead to the photoexcited molecules decaying non-radiatively very quickly. 

The PEPIPICO map between the time of flight of the faster ion (time of flight~$1$) and the slower ion (time of flight~$2$) is a useful representation to get a deeper understanding on the fragmentation dynamics. 
Figure~\ref{figure: Mass_spectra_coincidence_map}~\text{b}) and \text{c}) represent the PEPIPICO map for $287.9$~eV and $292.4$~eV\cite{sen_fragmentation_2022} photon energies, respectively. Fragment ions produced in the same fragmentation event appear as an island in the PEPIPICO map, and the dynamics of the fragmentation determine the slope and shape of the islands~\cite{eland_dynamics_1991}. The horizontal and vertical dashed lines represent the mass to charge ratios of the slower and faster ions, respectively, for the three fragmentation channels marked by red rectangular boxes, discussed later.
Significant instability of the dicationic molecules leads to many-body dissociation, forming more stable singly charged and/or neutral fragments. Broad islands can be seen in the PEPIPICO map, which suggests that after C~$1$\textit{s} ionization, there are multi-body dissociation processes that are accompanied by the release of neutral fragments. This could also result from overlapping fragmentation channels producing identical ionic fragments. \par
Further, Auger decay can remove multiple electrons from the system (double Auger decay or cascaded Auger decay); however, the probability of electron ejection decreases with the number of Auger electrons. According to the prediction model of Roos \textit{et al.}~\cite{roos_abundance_2018}, the empirical estimate of the triply ionized state for C~$1$\textit{s} ionization from all of the C atoms is between $10-16$ \%~\cite{sen_fragmentation_2022}. 
Following this assumption, it is assumed that triple ionization will have less impact on ion-ion coincidences, allowing us to focus on the doubly ionized state produced by normal Auger decay. 

The PEPIPICO maps look similar with different intensities, so the remaining discussion addresses the $292.4$~eV photon energy. From the PEPIPICO map, three islands are chosen for further analysis, designated by the red rectangular boxes in the PEPIPICO map. These islands are formed by masses $m/q$ of ($15$ \& $95$), ($29$ \& $81$), and ($42$ \& $95$). 
  
\begin{figure*}[ht]
\includegraphics[width=16 cm]{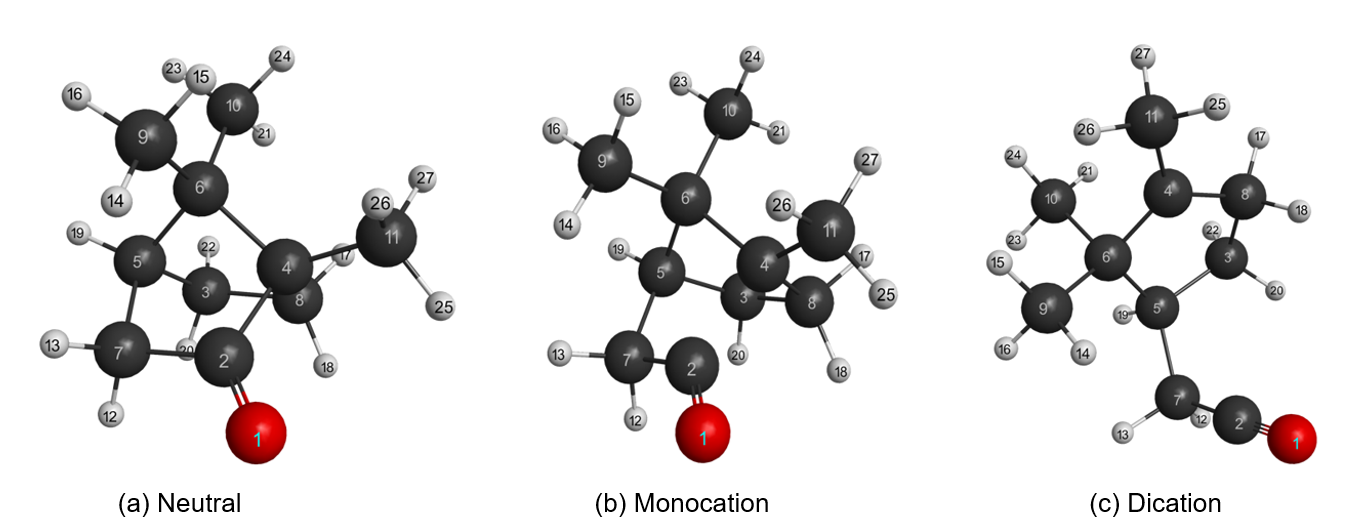}
\caption{Optimized structures for the camphor molecule in \text{a}) neutral\cite{sen_fragmentation_2022}, \text{b}) monocationic, and \text{c}) dicationic state obtained through DFT simulation.}
\label{figure: Optimized_structures} 
\end{figure*}
The optimized structures of the \text{a}) neutral\cite{sen_fragmentation_2022}, \text{b}) monocationic, and \text{c}) dicationic state of the camphor molecule are shown in Figure \ref{figure: Optimized_structures}. A DFT simulation of the exchange-correlation functional of type B$3$LYP utilising the $6311$-G basis set was used to optimize the structure. The density matrix's convergence threshold is maintained at $10^{-6}$ during optimization. To validate our geometry optimization, we compared it with the neutral ($1$\textit{S})-($-$)-Camphor's known geometry from previous studies~\cite{de_castilho_single_2014, brunelli_low-temperature_2002}. A detailed description can be found in our recent work~\cite{sen_fragmentation_2022}. Note that the bond orders shown in the figures are incorrect. The significant transformation in geometry during dication formation is quite intriguing in the case of the camphor molecule. Molecular structural rearrangement is also observed in acrylonitrile dication~\cite{Itälä_2009} during migrating of a terminal or middle-chain hydrogen (deuterium) to the cyano group. In another study, the formation of H$_3^+$ in methanol and ethanol~\cite{Ekanayake_2018} via the roaming mechanism of H$_2$, formed by the elongation of C–H bonds and the narrowing of the H–C–H angle on the $\alpha$ - carbon atom in their corresponding doubly charged ions, has been seen.

The Gibbs free energy of neutral, singly charged and doubly charged camphor molecules is $0$, $187.9$ and $446.1$~kcal/mol, respectively~\cite{de_castilho_single_2014}. This large difference in Gibbs free energy indicates the highly unstable nature of camphor dication, which rules out the presence of dicationic camphor molecule and its doubly charged ionic fragments in mass spectra. 
The non-observation of camphor dication may also be due to the availability of decay pathways where the camphor dictation produced due to the vertical excitation of the camphor molecule, can surpass the Coulomb barrier and therefore dissociates. Molecular fragmentation has also been found to be significantly influenced by ultrafast charge redistribution and ultrafast molecular restructure, as demonstrated by earlier studies~\cite{erk_ultrafast_2013, hishikawa_hydrogen_2004, xu_tracing_2009}. The stability of the dictation via geometry optimization and molecular fragmentation are both competitive processes following double ionization. For the camphor dication, the insufficient time for geometric optimization or charge redistribution likely results in relaxation via fragmentation.
\begin{figure*}[ht]
\includegraphics[width=17 cm]{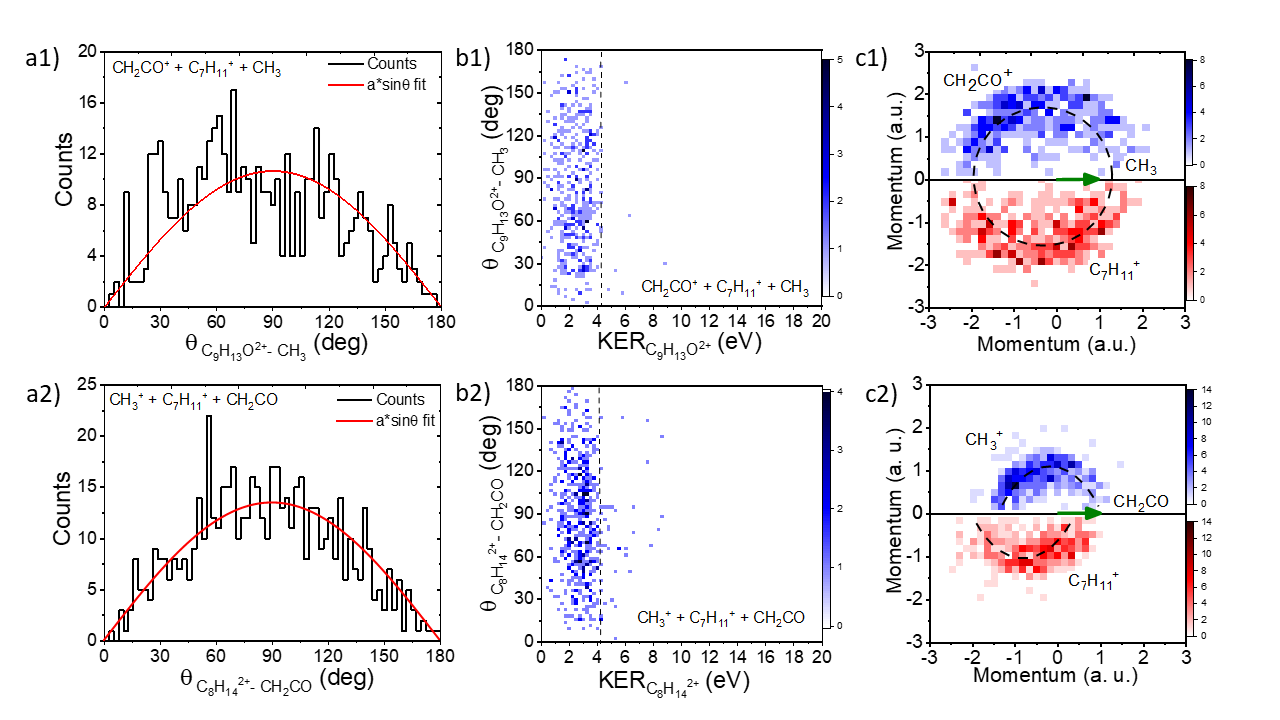}
\caption{The angular distribution of the $2$nd step with respect to the $1$st step of sequential dissociation [a1) and a2)], density plot between the angular distribution and the kinetic energy release in the $2$nd step of dissociation process [b1) and b2)] and the corresponding Newton diagram [c1) and c2)] are presented for the channels ($1$) C$_{10}$H$_{16}$O$^{2+}$ $\rightarrow$ CH$_2$CO$^+$  + C$_7$H$_{11}^+$  + CH$_3$ and ($2$) C$_{10}$H$_{16}$O$^{2+}$ $\rightarrow$ CH$_3^+$  + C$_7$H$_{11}^+$  + CH$_2$CO respectively. The momentum of the neutral fragment is set to 1.0 a.u. and plotted along the X-axis, represented by a green arrow. The neutral fragment momentum is used to normalize the momenta of the other two charged fragments, which are then plotted in the upper and lower halves with respect to the X-axis.}
\label{figure: Native_frame_1}
\end{figure*}  
To shed light on the instability of camphor dication, one can also compare the Wiberg bond order indices between other C-C bonds as mentioned in Castilho \emph{et al.}~\cite{de_castilho_single_2014}. 
The Wiberg bond index is known to be proportional to the bond strength; therefore, from the value of the Wiberg index, bond stability and the probability of bond breaking can be estimated.
The value of the Wiberg index of the C2-C4 bond in monocationic and dicationic states are 0.2128 and 0.004, respectively. These low values of Wiberg indices suggest that this bond has a high likelihood of rupture and is, therefore, broken in its monocationic state.
Comparatively small Wiberg indices at dicationic state for C$6$-C$10$ ($0.8517$), C$6$-C$9$ ($0.9905$), and C$5$-C$7$ ($0.9433$) indicate that the camphor molecule in its dicationic state can dissociate with a fragment CH$_3$ or CH$_2$CO.
  
\subsection{Three-body dissociation of dicationic camphor
molecule}
From the PEPIPICO, we chose three islands for further analysis in such a way that one of the fragments is either CH$_3$ ($m/q~15$) or CH$_2$CO ($m/q~42$) in a singly charged or neutral state. 
The release of the methyl group CH$_3$ ($m/q~15$) and ethenone group CH$_2$CO ($m/q~42$) from the camphor molecular ion has also been predicted by Weinberg \emph{et al.}~\cite{weinberg_mass_1966} and Dimmel \emph{et al.}~\cite{dimmel_1967_preferential} 
Again, due to detector dead time, our detector could not fully resolve the fragmentation channels having fragments with comparable masses. To avoid this, we consider only those channels that have fragments whose masses are largely separated from each other.
These three islands are predicted as channels for ($1$) CH$_2$CO$^+$ ($m/q~42$) + C$_7$H$_{11}^+$ ($m/q~95$) + CH$_3$ ($m/q~15$), ($2$)  CH$_3^+$ ($m/q~15$) + C$_7$H$_{11}$ ($m/q~95$) + CH$_2$CO ($m/q~42$) and ($3$) C$_2$H$_5^+$ ($m/q~29$) + C$_6$H$_9^+$ ($m/q~81$) + CH$_2$CO ($m/q~42$). Although (CHO$^+ + $C$_6$H$_9^+ + $C$_3$H$_6$) fragmentation channel is also possible, but it requires molecular rearrangement before fragmentation. Therefore, we do not consider these types of fragmentation channels. We only focussed on those channels where one of the fragment ions is either methyl (CH$_3$) or an ethenone group (CH$_2$CO).
A slope analysis of these islands from linear regression gives the approximate slopes of $-0.97 (\pm 0.03)$, $-1.33 (\pm 0.05)$, and $-0.71 (\pm 0.02)$, respectively. 
These values are close to the theoretically calculated value for deferred charge separation type of dissociation. In our previous report~\cite{sen_fragmentation_2022}, using molecular dynamic simulation, we showed that the dicationic camphor molecule dissociates by the deferred charge separation process. 
In this report, we have extended our previous work and experimentally shown three new dissociation channels, as discussed below: \\

\textbf{ \small
1. Fragmentation channel: C$_{10}$H$_{16}$O$^{2+}$ $\rightarrow$ CH$_2$CO$^+$  + C$_7$H$_{11}^+$  + CH$_3$
} \\
To discuss the fragmentation dynamics of these channels, we used the ``native frame'' method. We assume that the dissociation occurs sequentially in a two-step process, where CH$_3$ leaves in the first step and C$_9$H$_{13}$O$^{2+}$ is an intermediate ion. We calculate the correlated momentum at each step with respect to each center-of-mass reference frame. As the intermediate ion rotates after the first step of dissociation, the angular correlation between the first and second steps of dissociation is lost. Therefore, the angular distribution becomes isotropic in nature and follows a ``$\sin\theta$'' distribution (solid red line), as shown in Figure~\ref{figure: Native_frame_1}~a1). The isotropic character of this distribution can be elucidated by considering the rovibrational excitation of intermediate ions, resulting in a uniform angular distribution (over the sphere). This effect requires that the lifetime of the electronic states of C$_9$H$_{13}$O$^{2+}$ greatly exceeds its rotational time period.
A rotational period of $\sim 244$~ps is calculated~\cite{kisiel_structure_2003} for the neutral camphor molecule for $J = 1$ rotational state, but in our case, the dicationic intermediate ion is rotationally as well as vibrationally excited. Due to a lack of literature, it is difficult to comment on the lifetime and the rotational time period of the intermediate ion. In our previous study~\cite{sen_fragmentation_2022}, we have observed that following the emission of neutral CO, the intermediate ion rotates for $ \sim 150$~fs and $\sim 290$~fs before further fragmentation, for the two channels (\textbf{i}) CH$_3^+$ + C$_8$H$_{13}^+$ + CO, and (\textbf{ii}) C$_4$H$_7^+$ + C$_5$H$_9^+$ + CO, respectively. Now, in our experiments, the flight time of the detected ions is in $\mu$s, so the intermediate metastable ions are dissociating before reaching the detector.  Therefore, we mostly missed them.

The density plot between the angular distribution [as presented in Figure~\ref{figure: Native_frame_1}~a1)] and the kinetic energy release in the second step of dissociation is shown in Figure~\ref{figure: Native_frame_1}~b1). The uniform angular distribution limited to the dashed black line around $\sim 4.2$~eV in our case indicates that the process underlined in the channel discussed is a sequential type with the intermediate C$_9$H$_{13}$O$^{2+}$. The corresponding momenta correlation map or Newton diagram is shown in Figure~\ref{figure: Native_frame_1}~c1). The momentum of the neutral fragment, CH$_3$ is set to 1 a.u. and plotted along the X-axis. The momentum of the other two charged fragment ions CH$_2$CO$^+$ and C$_7$H$_{11}^+$ are normalized with respect to the momentum of the neutral fragment ion and plotted in the upper and lower half of the X-axis. Due to the rotation of the metastable intermediate ion before dissociating into two charged fragments, the angular distribution of the momenta of these charged fragments can take arbitrary values, resulting in a semi-circular feature on the Newton diagram. Thus, the semi-circular nature of this diagram is also a signature of sequential dissociation, which is also evidenced in our case by the corresponding Newton diagram, shown in Figure~\ref{figure: Native_frame_1}~c1).
\begin{figure*}[ht]
\includegraphics[width=17 cm]{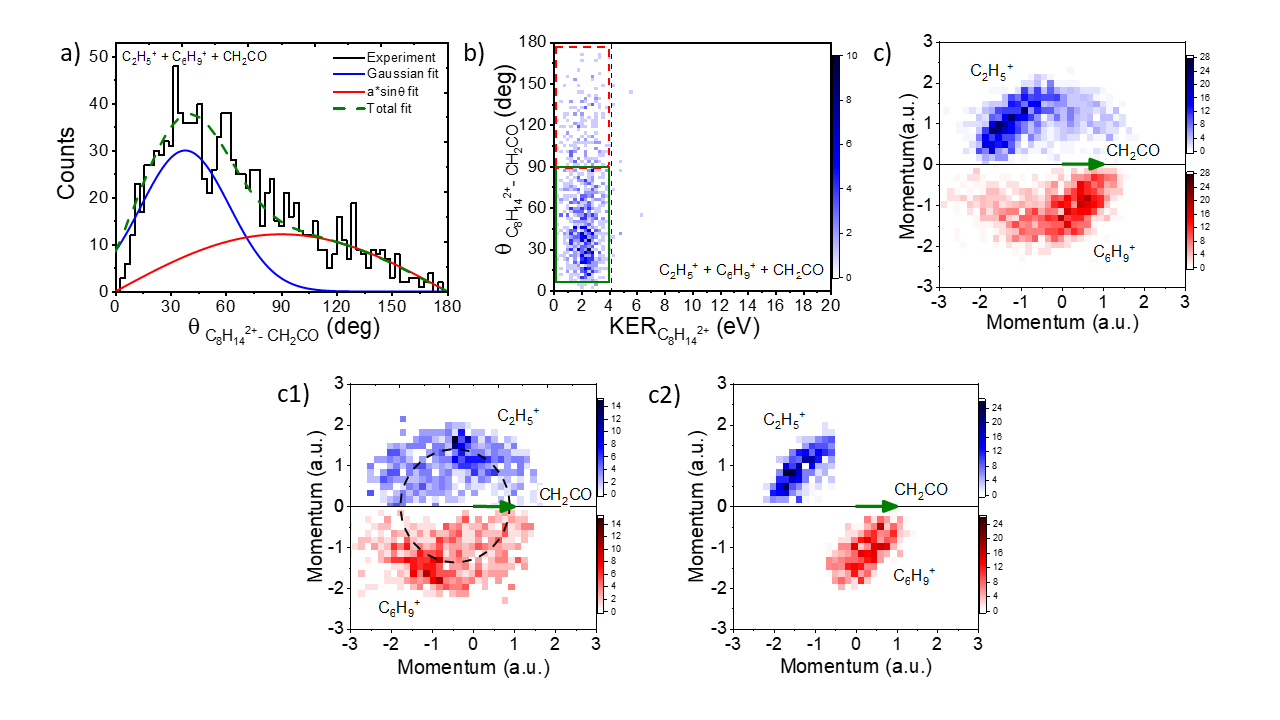}
\caption{The angular distribution of the $2$nd step of sequential dissociation [a)], density plot between the angular distribution and the kinetic energy release in the $2$nd step of dissociation process [b)], and the corresponding Newton diagram [c)] are presented for the channels ($3$) C$_{10}$H$_{16}$O$^{2+}$ $\rightarrow$ C$_2$H$_5^+$  + C$_6$H$_{9}^+$  + CH$_2$CO respectively. The events corresponding to sequential and concerted dissociation are extracted by applying the conditions shown by the red dashed rectangle and green solid rectangle in the density plot, respectively. The corresponding Newton diagrams are plotted in c1) and c2) for sequential and concerted dissociation, respectively. The momentum of the neutral fragment is set to 1.0 a.u. and plotted along the X-axis, represented by a green arrow. The neutral fragment momentum is used to normalize the momenta of the other two charged fragments, which are then plotted in the upper and lower halves with respect to the X-axis.}
\label{figure: Native_frame_2}
\end{figure*}
 \textbf{ \small
2. Fragmentation channel: C$_{10}$H$_{16}$O$^{2+}$ $\rightarrow$ CH$_3^+$  + C$_7$H$_{11}^+$  + CH$_2$CO} \\
Although the statistics for this channel are low, we can still get some information from native frame analysis. The solid red line in Figure~\ref{figure: Native_frame_1}~a2) represents the nearly perfectly fit ``$\sin\theta$'' distribution and indicates the existence of a purely sequential type decay process with an intermediate C$_{8}$H$_{14}^{2+}$ ion. The uniform distribution in the density plot of $\theta_{\text{C}_{8}\text{H}_{14}^{2+}, \text{CH}_2\text{CO}}$ and KER$_{\text{C}_{8}\text{H}_{14}^{2+}}$ in Figure~\ref{figure: Native_frame_1}~b2) confirms this conclusion. The Newton diagram for this channel is shown in Figure~\ref{figure: Native_frame_1}~c2), where a neutral CH$_2$CO is released in the first step, and then the intermediate C$_{8}$H$_{14}^{2+}$ rotates for a significant time that could exceed the rotational time period. In the final step, the intermediate dissociates into two charged fragments CH$_3^+$ and C$_7$H$_{11}^+$, in an almost back-to-back emission. The experimentally measured angular distribution between these two charge fragments peaked around $\sim 160^{\circ}$. The semi-circular nature of the Newton diagram also suggests that the dominating process is sequential type.

Due to low statistics, the presence of a concerted decay with negligible counts cannot be distinguished from the dominating sequential decay in both channels (1) CH$_2$CO$^+$  + C$_7$H$_{11}^+$  + CH$_3$, and (2) CH$_3^+$  + C$_7$H$_{11}^+$  + CH$_2$CO, discussed above. Considering the initial charge separation decay for these two channels with the intermediate ion C$_8$H$_{14}^+$ and C$_9$H$_{13}$O$^+$, respectively, we observed scattered distributions in the density plot [figures are not shown] between the angular distribution and the kinetic energy release in the $2$nd step of dissociation. This suggests that the initial charge separation decay process is imperceptible for these channels.

 \textbf{ \small
3. Fragmentation channel: C$_{10}$H$_{16}$O$^{2+}$ $\rightarrow$ C$_2$H$_5^+$  + C$_6$H$_{9}^+$  + CH$_2$CO} \\

Figure~\ref{figure: Native_frame_2}~a) shows the angular distribution between the $2$nd step and $1$st step of dissociation for this channel. We observe a clear maximum around $\sim 
30 ^{\circ} - 50 ^{\circ}$ superimposed with the isotropic distribution. We fitted the total distribution with a Gaussian distribution (for the localized dense region) and a ``$\sin\theta$'' distribution (for the isotropic region), represented by the solid blue and red lines, respectively. The Gaussian distribution almost diminishes around $\theta_{\text{C}_{8}\text{H}_{14}^{2+}, \text{CH}_2\text{CO}} = 90 ^{\circ}$, therefore we consider that the distribution for $\theta_{\text{C}_{8}\text{H}_{14}^{2+}, \text{CH}_2\text{CO}} =
90 ^{\circ} - 180 ^{\circ}$ is isotropic, and for $\theta_{\text{C}_{8}\text{H}_{14}^{2+}, \text{CH}_2\text{CO}} =
0 ^{\circ} - 90 ^{\circ}$, the two distributions overlap with each other. This highly intense region is also visible in the density plot between $\theta_{\text{C}_{8}\text{H}_{14}^{2+}, \text{CH}_2\text{CO}}$ and KER$_{\text{C}_{8}\text{H}_{14}^{2+}}$, overlapping the uniform distribution, as shown a solid green rectangle in Figure~\ref{figure: Native_frame_2}~b). The uniform distribution is a signature of sequential decay, whereas the localized intense distribution clearly indicates the presence of a concerted decay process.
The Newton diagram, shown in Figure~\ref{figure: Native_frame_2}~c), for this channel consists of a localized dense region overlapping with a uniform distribution. 

To separate these two features in the Newton diagram, we have considered that the feature for $\theta_{\text{C}_{8}\text{H}_{14}^{2+}, \text{CH}_2\text{CO}} = 90 ^{\circ} - 180 ^{\circ}$ is isotropic (as shown in Figure~\ref{figure: Native_frame_2}~a)). In addition, the other half of the isotropic distribution for $\theta_{\text{C}_{8}\text{H}_{14}^{2+}, \text{CH}_2\text{CO}} = 0 ^{\circ} - 90 ^{\circ}$ is generated by generating random numbers (integers) while maintaining equal total counts for both halves. We also made sure that the maximum number generated in the other half should not exceed the maximum number in the known isotropic distribution part. In this way, we have constructed the complete Newton diagram for isotropic distribution, shown in Figure~\ref{figure: Native_frame_2}~c1). The semi-circular nature of the Newton diagram also reveals that the underlying mechanism is sequential. To obtain the Newton diagram for localized regions, we subtracted the Newton diagram for isotropic distribution from the total Newton diagram and the result is shown in Figure~\ref{figure: Native_frame_2}~c2).
The Newton diagram clearly indicates a back-to-back emission of C$_2$H$_5^+$ and C$_6$H$_9^+$ ions, leaving the neutral fragment with almost negligible momentum. Furthermore, the kinetic energy released in the second step of the initial charge separation dissociation process with the intermediate C$_{8}$H$_{11}$O$^{+}$ for the channel (3) is low, and therefore examining this channel is difficult.

Note, for all three deferred charge separation channels, it is observed in the Newton diagram that the radius of the semi-circular distribution decreases with the increase of the ratio of masses between the larger fragment and the smaller fragment in the $2$nd step of the decay processes as observed in our previous study.~\cite{sen_fragmentation_2022} \par

\subsection{Total kinetic energy release and fragment ion kinetic energy spectra}
\begin{figure}
\includegraphics[width=8 cm]{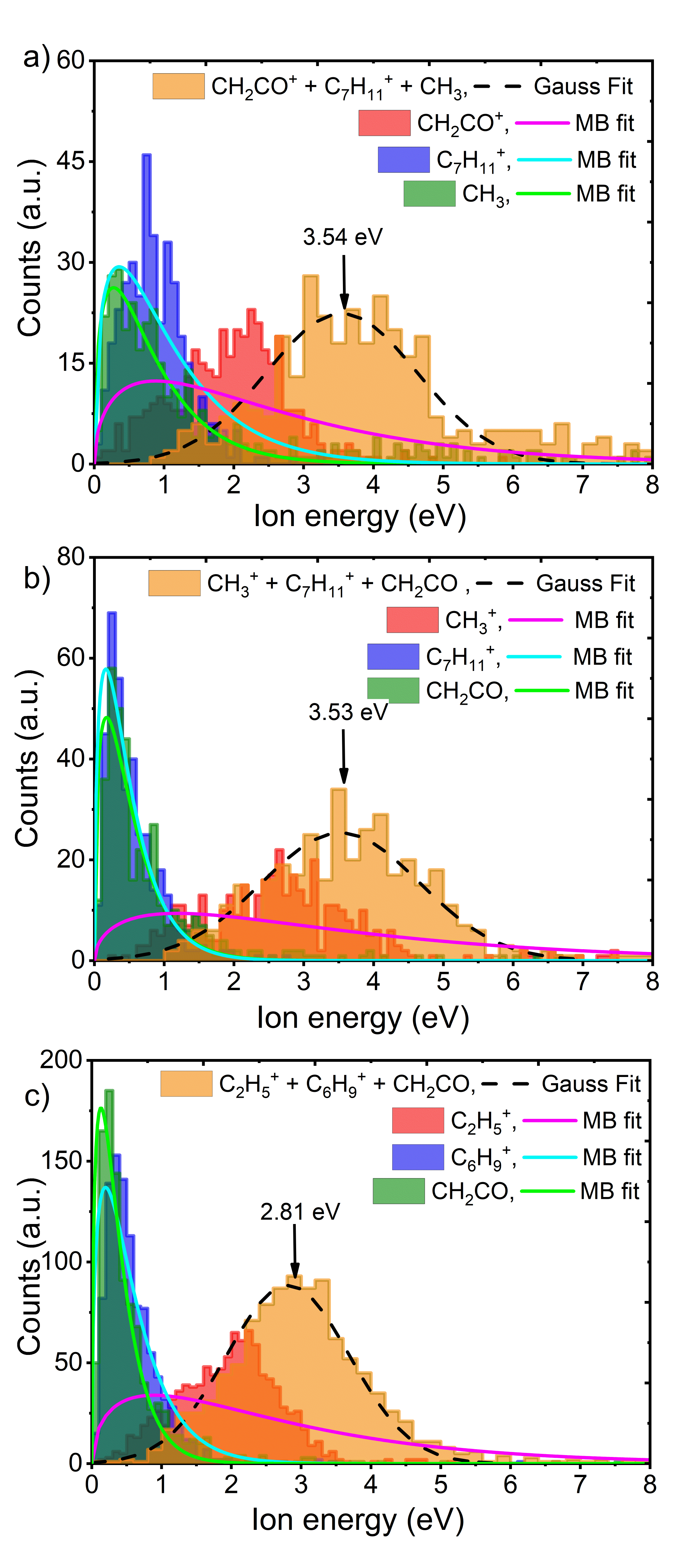}
\caption
{The KER distributions and kinetic energies (KEs) of the individual fragment ions for the three-body fragmentation of C$_{10}$H$_{16}$O$^{2+}$ to the channels ($1$) CH$_2$CO$^+$ + C$_7$H$_{11}^+$ + CH$_3$, ($2$) CH$_3^+$ + C$_7$H$_{11}^+$ + CH$_2$CO, and ($3$) C$_2$H$_5^+$ + C$_6$H$_9^+$ + CH$_2$CO are shown in filled colours histograms. The Gaussian fit for the total KER distribution is depicted by the black dashed lines, whereas the Maxwell-Boltzmann distributions are fit to the fragment ion kinetic energy spectra and are shown by solid lines with different colours for different fragment ions.
}
\label{figure: Ion_energy}
\end{figure}  
The dissociation of the camphor molecule post C~$1$\textit{s} ionization can occur via Auger decay, where the initial core hole can decay on a femtosecond time scale~\cite{carroll_carbon_2002, Carrol_core_hole_2000}, and this leaves the molecule in a dicationic state with a significant amount of energy. The key factor controlling the fragmentation pattern is the quantity of energy transferred to the molecular ion after ionization, and, therefore, the fragmentation pattern should be unaffected by the excitation/ionization process involved. This excess energy is allocated across all internal degrees of freedom~\cite{erdmann_general_2021}. Even in the case of resonant Auger decay, internal energy has been found to be the crucial parameter affecting the fragmentation pattern, as observed in a recent study involving N-methyltrifluoroacetamide~\cite{salen_resonant_2020}. Kukk \emph{et al.}~\cite{kukk_internal_2015} also reported the high sensitivity of branching ratios on molecular internal energies for the case of the fragmentation of thiophene, where taking into account the internal energy (temperature) as a parameter produced an excellent agreement with the experiments, which suggested that internal energies at all molecular sites would redistribute very quickly~\cite{calegari_ultrafast_2014, Remacle_Ultrafast_2006}. The initial ionization site and the fragmentation pattern were clearly related in a different study of the fragmentation of chloro- and bromo-acetic acid after C 1s ionization. This was corroborated by comparing the internal energies after site-specific Auger decays~\cite{levola_ionization-site_2015}. \par

The KER distribution and the individual ionic fragment kinetic energy distribution for each fragmentation channel discussed above are shown in Figure~\ref{figure: Ion_energy}. We fitted a Maxwell-Boltzmann (MB) distribution to the fragment ion kinetic energy spectra, whereas we used a Gaussian distribution to fit the total kinetic energy release to compute the peak energy. It is observed that the KER released by channels ($1$) and ($2$) peaked around $\sim 3.5$~ eV, thus indicating that the energetically close states are accessed by the camphor molecules, post C~$1$\textit{s} ionization. As the release of the neutral fragment occurs in the $1$st step of the deferred charge separation channels ($1$) and ($2$) in a non-Coulombic way, the kinetic energies carried by the neutral fragments are low, which are depicted by the light green filled areas in Figure~\ref{figure: Ion_energy}~a) and b). The energy carried away by CH$_3$ for the channel ($1$) has a maximum near $\sim 0.55$~eV, whereas the kinetic energy carried away by the neutral fragment CH$_2$CO in the channel ($2$) peaked at $\sim 0.34$~eV. Thus, the momentum acquired by CH$_3$ fragment in the first step of dissociation in the channel ($1$) is higher and consequently, the corresponding intermediate ion gains more momentum than that of CH$_2$CO fragment in the channel ($2$). This is also reflected in the corresponding Newton diagram for the channel ($2$), as a smaller semi-circular distribution, as shown in Figure~\ref{figure: Native_frame_1}~c2), compared to the Newton diagram for the channel ($1$) as shown in Figure~\ref{figure: Native_frame_1}~c1). The KER for channel ($3$) is lower than for the other two channels and peaked at $\sim 2.8$~eV. The lower KER for this channel may result from the relaxation of the dicationic state to a lower excited state before fragmentation. However, the possibility of being excited into different states via conical intersections may also be possible and cannot be ignored in such a system where the excited states are densely packed.
\section{Conclusions}
We studied the ionic fragmentation of the dicationic camphor molecules at photon energies of $287.9$~eV (below C $1$\textit{s} ionization potential of camphor molecule) and $292.4$~eV (above the C~$1$\textit{s} ionization potential for skeletal C) using VMI-PEPIPICO techniques. Ion mass spectra and the ToF coincidence maps reveal the presence of the same ionic fragments at both photon energies, indicating that the fragmentation depends on the final dicationic states rather than the initial method of excitation. \par
We report three new fragmentation channels of dicationic camphor molecules: three deferred charge separation channels ($1$) CH$_2$CO$^+$ + C$_7$H$_{11}^+$ + CH$_3$, ($2$) CH$_3^+$ + C$_7$H$_{11}^+$ + CH$_2$CO, and ($3$) C$_2$H$_5^+$ + C$_6$H$_9^+$ + CH$_2$CO, along with one concerted decay channel C$_2$H$_5^+$ + C$_6$H$_9^+$ + CH$_2$CO, for the first time. From the geometry optimization of the camphor dication and the Wiberg bond indices values from Castilho \emph{ et al.}~\cite{de_castilho_single_2014}, we have inferred which bonds have a high probability of rupture. With the help of the native frame analysis method and the Newton diagram, we observed the dominating nature of sequential dissociation.

This study motivates future experiments on studying the fragmentation dynamics of large molecules by the PEPIPICO technique and photoelectron spectroscopy. Observing the photoelectron and Auger electron spectra correlated to a specific fragmentation channel can reveal intricate details about the fragmentation dynamics. This study also sets the base for time-resolved studies, where a deeper understanding of the real-time evolution of different fragmentation channels can be achieved by using state-of-the-art facilities like FELs.
\section*{Acknowledgments}
S. R. K. and S. D. are grateful to the Department of Science and Technology (DST), India, and ICTP, Trieste, who supported this campaign ($\#$20185050) to carry out this campaign at the Elettra Synchrotron facility. 
S. D. gratefully acknowledges support from the Indo-Elettra program. S. R. K. acknowledges support through the Indo-French Center for Promotion of Academic Research (CEFIPRA) and the DST-DAAD bilateral research program of the Ministry of Science and Technology, Govt. of India, with Deutsche Akademischer Austauschdienst, Germany. 
S. R. K. thanks DST and the Max Planck Society for financial support. S. R. K., V. S. and S. D. acknowledge funding from the SPARC program, MHRD, India. S. R. K. and S. D. gratefully acknowledge financial support from IIT Madras through the Quantum Center of Excellence for Diamond and Emergent Materials (QuCenDiEM) group as part of the Institute of Eminence (IoE) program (project $\#$SB20210813PHMHRD002720).
V. S. thanks DST-SERB and DAE-BRNS, and V. S., R. G. and S. S. thank IMPRINT ($\#$5627) for financial support.
M. M. and L. B. L. acknowledge financial support by the Danish Council for Independent Research Fund (DFF) via Grant No. 1026-00299B and by the Carlsberg Foundation. We thank the Danish Agency for Science, Technology, and Innovation for funding the instrument center DanScatt. L. B. L. acknowledges financial support by the Villum foundation via the Villum Experiment grant No. 58859. 
\section*{References}
\bibliography{references} 
\bibliographystyle{unsrt}
\end{document}